# Magnetic Correlation of Na$_x$CoO$_2$ and Successive Phase Transitions of Na$_{0.5}$CoO$_2$ -NMR and Neutron Diffraction Studies-

Mai Yokoi, Taketo Moyoshi, Yoshiaki Kobayashi, Minoru Soda, Yukio Yasui and Masatoshi Sato*

*Department of Physics, Division of Material Science, Nagoya University, Furo-cho, Chikusa-ku, Nagoya 464-8602*

and

Kazuhisa Kakurai

*Advanced Science Research Center, Japan Atomic Energy Research Institute, Tokai-mura, Naka-gun, Ibaraki 319-1195*

Data taken by various kinds of methods including NMR/NQR and neutron scattering are presented for Na$_x$CoO$_2$, in particular, for Na$_{0.5}$CoO$_2$. Attentions have also been paid to the $x$ dependence of the electronic nature. The pseudo-gap-like behavior observed in the region of $x<x_c\sim 0.6$ is emphasized. For samples with $x\leq 0.6$, signals from two kinds of Co sites with different quadrupole frequencies $\nu_Q$, ~4.1 MHz and ~2.8 MHz were observed, and the former value of $\nu_Q$, which is nearly equal to that of the superconducting system Na$_{0.3}$CoO$_2\cdot y$H$_2$O, becomes dominant with decreasing $x$ to 0.3. For Na$_{0.5}$CoO$_2$, the Co sites with the larger $\nu_Q$ value have the larger magnetic moments. They align antiferromagnetically at $T_{c1}\sim 87$ K with their direction within the plane, while the Co sites with the smaller $\nu_Q$ value have the smaller moments. They align along the direction parallel to the $c$ axis. The ordered moments at the two distinct sites exhibit the same $T$ dependences, indicating that the existence of these sites is not due to a macroscopic phase separation. Although we have observed the splitting of the zero-field NMR signals from the Co sites with the smaller $\nu_Q$ value at the second transition temperature $T_{c2}\sim 53$ K, any other anomaly of the ordered magnetization has not been observed there in the $T$ dependences of the neutron and the NMR data. As $x$ decreases, the two-dimensional nature of the electrons is enhanced and the antiferromagnetic correlation increases. Based on these results, the superconducting pair state in Na$_{0.3}$CoO$_2\cdot y$H$_2$O is discussed.

KEYWORDS: Na$_x$CoO$_2$, NMR, Neutron Diffraction, Superconductivity

## 1. Introduction

Na$_x$CoO$_2$ has a stacking of CoO$_2$ layers with Na atoms between them.[1] By de-intercalating the Na atoms from the system and then intercalating the H$_2$O molecules between the layers, Na$_{0.3}$CoO$_2\cdot y$H$_2$O ($y\sim 1.3$) with the superconducting transition temperature $T_c\sim 4.5$ K can be obtained.[2] Because the layers consist of the triangular lattice of Co atoms, strong magnetic fluctuations caused by the possible geometrical frustration, may play an important role in realizing the superconductivity, as was proposed in high-$T_c$ Cu oxides.[3] Another interesting point of view on the superconductivity stems from the fact that Co$^{3+}$ and/or Co$^{4+}$ ions in various Co oxides often exhibit a spin state change:[4-6] The low-spin (LS, $t_{2g}^6$) ground state of the 3$d$ electrons of Co$^{3+}$ in the pseudo cubic crystal field, for example, often changes to the intermediate-spin (IS, $t_{2g}^5 e_g^1$) or high-spin (HS, $t_{2g}^4 e_g^2$) state with varying temperature $T$ or with the substitution of other constituent elements. Because the $e_g$-electrons of the ions in the IS or HS state induce the double exchange interaction between spins in the $t_{2g}$ orbitals, the metallic and ferromagnetic (or nearly ferromagnetic) state is often realized in Co oxides. It may open a way to realize an unconventional superconductivity induced by inter band interactions near the ferromagnetic state. From these points of view, it is interesting to study the symmetry and the spin state of the superconducting pairs of Na$_{0.3}$CoO$_2\cdot y$H$_2$O. Actually, many theoreticians[7-10] have pointed out the possibility of the triplet pair state, though the inter band interaction does not necessarily induce the triplet pairing.[11]

Experimental studies on the superconducting pair state of Na$_{0.3}$CoO$_2\cdot y$H$_2$O have also been published by many authors: The possibility of the unconventional nature of the superconductivity have been suggested by observing the $T$ dependence of the electronic specific heat $C_{el}$.[12-18] However, one has to be careful to extract this conclusion, because there exist various problems, for example, the possible in-homogeneity of the superconducting $T_c$, which may arise from the change of the number of the H$_2$O molecules during the experimental processes. Many other ambiguous points have been pointed out by a recent report on $C_{el}$.[18] The NQR longitudinal relaxation rate ($1/T_1$) does not have the so-called coherence peak.[19-22] The non-existence of the peak is usually expected for anisotropic order parameter. The spin component of the NMR Knight shifts has been found to be suppressed in the superconducting phase for both directions of the applied magnetic fields $H$ within and perpendicular to the CoO$_2$ plane, as reported by our preceding papers,[22,23] indicating that the pairs are in the singlet state. The $T_c$-suppression rate by non-magnetic impurities has been found to be rather small,[24] which suggests that the superconducting order parameter is isotropic. Now, it seems to be necessary to explain these observations in a consistent way.

We have been studying, besides the studies on the superconducting phase of Na$_{0.3}$CoO$_2\cdot y$H$_2$O, physical properties of mother system $Na_xCoO_2$, by NMR/NQR, neutron scattering as well as other kinds of measurements of macroscopic quantities, to clarify the electronic nature of the $CoO_2$ planes. In the work, we have paid much attention to the dimensionality of the electron system and the $x$ dependence of the electronic properties. As for the dimensionality, recent neutron inelastic scattering experiments of magnetic excitations of $Na_{0.82}CoO_2$[25] and $Na_{0.75}CoO_2$[26, 27] have revealed the three dimensional nature of the magnetic interaction of the Co moments, which is somewhat unexpected from the structural characteristic of the system. How about the $x$ dependence of the dimensionality? It seems to be one of key questions in investigating whether the magnetic interaction has an important role for the electron pairing or not. For $Na_{0.5}CoO_2$, it is well known that there exist two transitions at $T_{c1} \sim 87$ K and $T_{c2} \sim 53$ K.[28] The understanding of the electronic and magnetic states of each phase of $Na_{0.5}CoO_2$ may present detailed information on various material parameters necessary for the identification of the origin of the superconductivity. Based on these ideas, we have studied magnetic properties of $Na_xCoO_2$, in particular $Na_{0.5}CoO_2$.

---

* corresponding author: (e-mail: e43247a@nucc.cc.nagoya-u.ac.jp)

In the present paper, we discuss the variation of the electronic nature of $Na_xCoO_2$, where the pseudo-gap-like behavior can be evidently found in the $T$ dependences of the magnetic susceptibility $\chi$ and the nuclear longitudinal relaxation rate $1/T_1$. We also report the experimental evidences for the magnetic ordering obtained for powder and single crystal samples of $Na_{0.5}CoO_2$ by NMR/NQR and neutron diffraction studies in both $T$ regions below $T_{c2}$ and between $T_{c1}$ and $T_{c2}$ and the possible magnetic structures are presented. Finally, discussion on the superconducting pair state is also presented in relation to the results obtained in this work.

**2. Experiments**

First, polycrystalline pellets of $Na_xCoO_2$ with $x \sim 0.75$ were prepared as described in ref. 29. The obtained samples were found to be the single phase. Then, these pellets or pulverized specimens were immersed into the $Br_2/CH_3CN$ solution for several days to de-intercalate Na atoms, where the $x$ value was controlled by controlling the amount of the solute $Br_2$. The obtained samples were washed with $CH_3CN$.

Single crystals of $Na_xCoO_2$ with $x \sim 0.72$ were grown by the floating zone (FZ) method, as reported in ref. 30. For these crystals, the Na de-intercalation was carried out as in the case of the polycrystalline pellets or pulverized specimens, where the de-intercalation could be achieved for several days. Then, the crystals were washed with $CH_3CN$. (We found it rather easy to de-intercalate Na atoms even for large crystal samples, though a long time is required to make the Na concentration homogeneous. However, the intercalation of $H_2O$ molecules is much more difficult than the Na de-intercalation.) The $x$ values of the samples were estimated by measuring the lattice parameter $c$ determined by X-ray diffraction. The reliability of the $x$ values can be confirmed from the fact that the two transitions, which are well known to take place for samples with $x=0.5$, are actually found at the $x$ value. To prepare aligned crystal samples for the NMR, the Na de-intercalated large crystal were crushed into platelets with typical sizes of $3 \times 2 \times 0.2$ mm$^3$, and the $c$ axes were aligned.

The magnetic susceptibilities $\chi$ were measured by using a Quantum Design SQUID magnetometer in the temperature range between 5 K and 450 K. The specific heats $C$ were measured by a thermal relaxation method in the temperature range between 2.0 K and 60 K. For $x=0.5$, the range was between 2.0 K and 100K. The electrical resistivity $\rho$ of the sample with $x=0.5$ was measured for a polycrystalline pellet by using the four terminal method.

The NMR measurements with nonzero applied field $H$, were carried out by a standard coherent pulsed NMR method. The spectra were taken by recording the spin echo intensity with the applied field being changed stepwise. The NQR measurements and the zero-field NMR in the magnetically ordered phase were also carried out, where the frequency was changed stepwise, too. The nuclear longitudinal relaxation rate $1/T_1$ of $^{59}Co$ nuclei was measured by measuring the $^{59}Co$ nuclear magnetization $m$, which is proportional to the spin-echo intensity, as a function of the time $t$ elapsed after the saturation pulses were applied.

Neutron measurements on $Na_{0.5}CoO_2$ crystals were carried out by using the spectrometers TAS-2 and T1-1 with the double axis-condition, both of which are installed at the thermal guide of JRR-3M of JAERI in Tokai. At TAS-2, the horizontal collimations were 17'(effective)-40'-80' and the neutron wavelength $\lambda$ was ~2.359 Å and at T1-1, the horizontal collimations were 14'(effective)-40'-60' and $\lambda$ was ~2.462 Å. Different crystals were set in Al-cans filled with exchange He gas. One of the crystals was used at TAS-2 with the [001] and [100] axes and the [001] and [110] axes in the scattering plane. Another one was used at T1-1 with the [001] and [110] axes in the scattering plane. The temperature was controlled by using the Displex type refrigerators. At TAS-2, the 002 reflection of Pyrolytic graphite (PG) was used for the monochromator and PG filters were placed in front of the second collimator and after the sample to eliminate the higher order contamination. At T1-1, they were put after the second collimator and after the sample. In the analyses of the data, the isotropic magnetic form factor for Co atoms was used. The absorption corrections were made.

**3. Experimental Results**
*3.1 Magnetic susceptibility and specific heat*

Figures 1(a) and 1(b) show the magnetic susceptibilities $\chi$ of the $Na_xCoO_2$ samples with various values of $x$ against $T$. In the former figure, data taken for polycrystalline or powder samples are shown together with that of the hydrated superconductor $Na_{0.3}CoO_2 \cdot 1.3H_2O$, while the latter shows data of single crystal samples taken for the applied magnetic field parallel to the $ab$-planes. For the single crystal sample with $x \sim 0.31$, $\chi$ is larger than the value obtained by the extrapolation of the $\chi$–$x$ curve from the region of $x > 0.31$. It is probably because the sample contains CoO (about 3 % of total weight) as an impurity phase. For the single crystal sample with $x \sim 0.75$, the $\chi$–$T$ curve exhibits an anomaly at ~22 K, indicating the existence of the antiferromagnetic (AF) transition.[31, 32] Although such a transition was not found for the polycrystalline sample with $x \sim 0.75$ shown in Fig. 1(a), we have often observed the antiferromagnetic transition even for other polycrystalline samples.

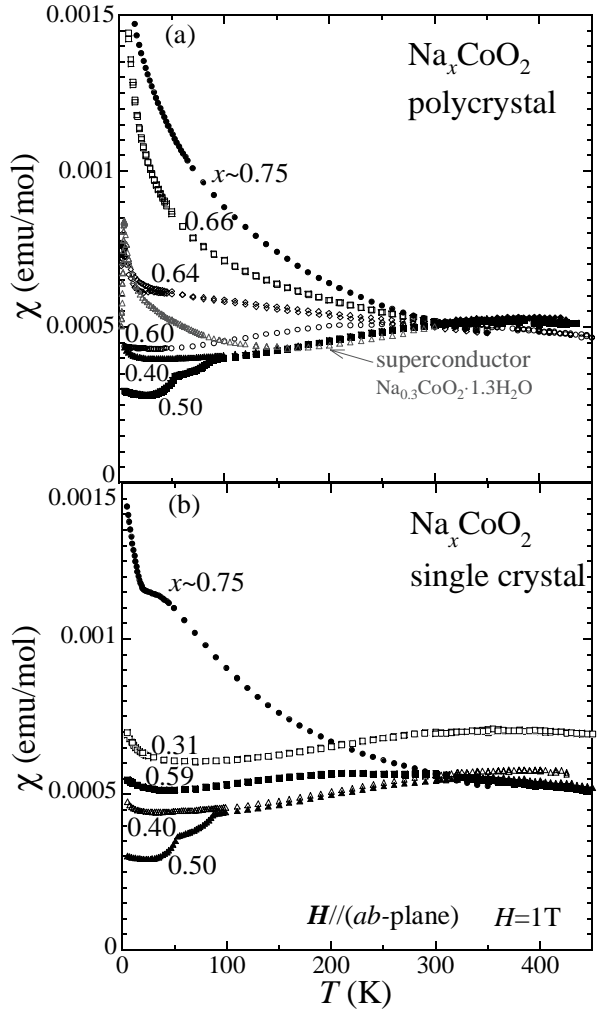

Fig. 1 Magnetic susceptibilities $\chi$ of polycrystalline samples (a) and single crystal samples ($H$//$ab$-plane) (b) of $Na_xCoO_2$ are shown for various values of $x$. Data of a hydrated sample of $Na_{0.3}CoO_2 \cdot 1.3H_2O$ is also shown in (a).

The characteristic of the $\chi$–$T$ curves in the region of $x \leq 0.6$ is rather different from that for $x \geq 0.6$, suggesting that the electronic state of $Na_xCoO_2$ can be divided by the boundary at $x = x_c \sim 0.6$: The $T$-dependence of $\chi$ observed for $x = 0.75$ is, roughly speaking, Curie-Weiss-like. In the region of $0.6 < x < \sim 0.7$, the $T$ dependence is weakened as $x$ decreases. In the region $x \leq x_c$, $\chi$ decreases with decreasing $T$, or the pseudo-gap-like behavior develops.

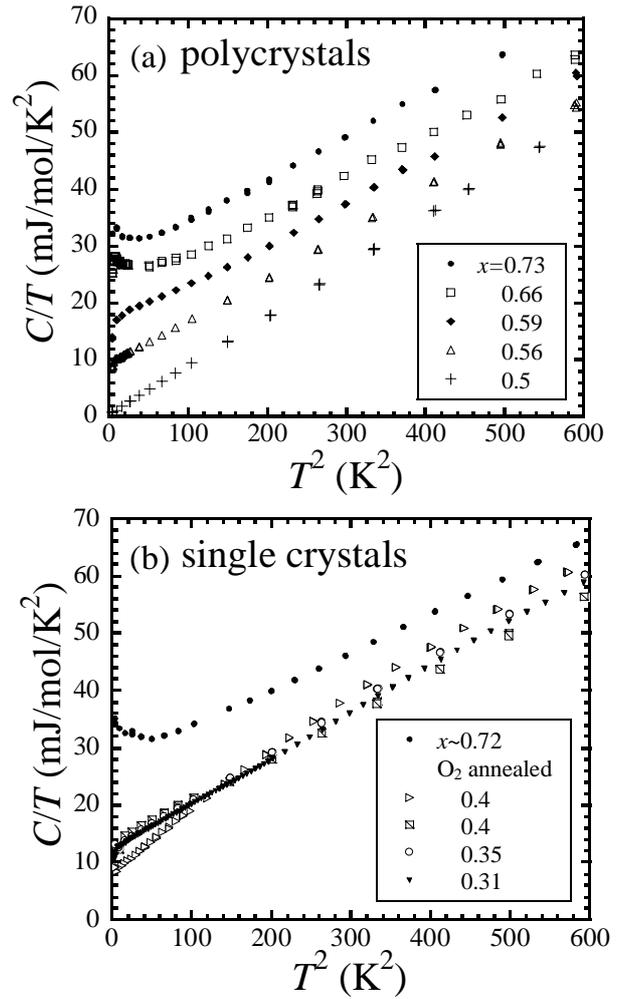

Fig. 2 Specific heats $C$ of polycrystalline samples (a) and single crystal samples (b) of $Na_xCoO_2$ are shown in the form of $C/T$-$T^2$ for various values of $x$.

The high temperature behavior observed for the sample with $x=0.5$ is similar to that of other samples with $x \leq x_c$. However, it exhibits two anomalies of $\chi$ at $T_{c1} \sim 87$ K and $T_{c2} \sim 53$ K, indicating two phase transitions.[28] It should be noted here that the existence of the boundary line was first pointed out by Foo et al.,[28] where the $x_c$ value was 0.5. We emphasize that in the region of $x < x_c \sim 0.6$, the magnetic susceptibility $\chi$ shows the existence of the pseudo gap in the present system similar to that of high $T_c$ Cu oxides. The existence of the pseudo gap is supported by the observation of the $T$ dependence of the $^{59}$Co nuclear relaxation rate $1/T_1$, as shown later.

In Figs. 2(a) and 2(b), data of the specific heats $C$ are shown in the form of $C/T$-$T^2$ for polycrystalline and single crystal samples, respectively. The phonon terms ($C/T \propto \beta T^2$) are roughly $x$-independent. In Fig. 3, the electronic specific heat coefficient $\gamma$ is plotted against $x$, where we can find again two distinct $x$ dependences of $\gamma$ roughly divided by the boundary at $x = x_c$. For $Na_{0.5}CoO_2$, $\gamma \cong 0$ is found, which is due to the metal-insulator transition at 53K. The enhancement of $C/T$ observed for $x > x_c$ at low temperatures would be due to the magnetic moment near the impurities or defects in the $CoO_2$

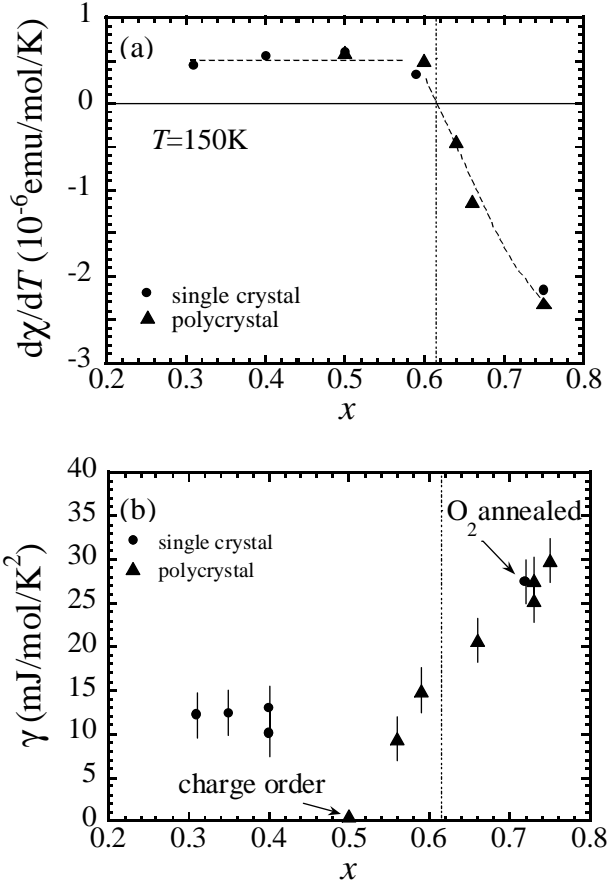

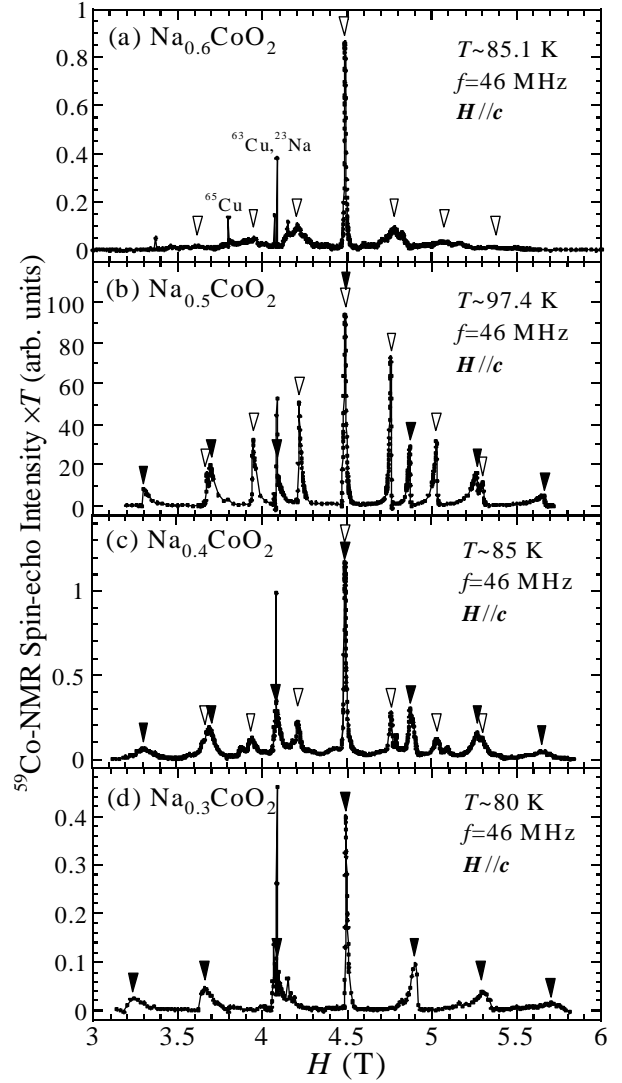

Fig. 3  (a) Temperature derivative of the magnetic susceptibility of $Na_xCoO_2$ determined at 150 K is plotted as a function of $x$, where the broken line is guide for the eyes. (b) Electronic specific heat coefficient γ of $Na_xCoO_2$ estimated from the linear part of $C/T$-$T^2$ curves are shown as a function of $x$.

Fig. 4  NMR spectra of $c$-axis aligned crystals of $Na_xCoO_2$ obtained with **H**//**c** are shown for $x$~0.6, 0.5, 0.4 and 0.3. The open symbols and closed symbols indicate the $^{59}$Co sites of $\nu_Q$~2.8 MHz and 4.1 MHz, respectively.

planes.

The temperature derivatives d$\chi$/d$T$ estimated at 150 K for the $\chi$-$T$ curves shown in Figs. 1(a) and (b), are plotted as a function of $x$ in Fig. 3 (a). In the region of $x$<0.62, d$\chi$/d$T$ is almost $x$-independent and positive, indicating that the pseudo gap-like behavior of the curves exists in this $x$ region. With increasing $x$, d$\chi$/d$T$ crosses zero at $x$=0.62 and with further increasing $x$ its absolute magnitude increases with increasing $x$, indicating that the Curie-Weiss-like behavior of $\chi$ is enhanced with increasing $x$. All these data indicate that there exists a clear boundary at $x$=$x_c$~0.62, which divides the $Na_xCoO_2$ system by their different characteristics of the electronic properties.

*3.2 NMR/NQR studies*

Figures 4(a)-4(d) show the NMR spectra observed with the applied magnetic field **H** parallel to the $c$-axis for aligned crystal samples with $x$~0.6, 0.5, 0.4 and 0.3, respectively, at temperatures indicated in the figures. (Note that $T$>$T_{c1}$ for $x$=0.5.) The NMR frequency $f$ was 46 MHz. The spectra for $x$~0.4 and 0.5 consist of two sets of signals of $^{59}$Co split by the $eqQ$ interaction with the quadrupole frequency $\nu_Q$~2.8 MHz (open symbols) and ~4.1 MHz (solid symbols), while the spectra observed for $x$~0.6 and 0.3 have only one set of signals split by the interaction with $\nu_Q$~2.8 MHz and $\nu_Q$~4.1 MHz, respectively. The fraction of the signal intensity corresponding to $\nu_Q$~2.8 MHz increases with increasing $x$ in the region $x$≤0.6. In the region of $x$>0.6, a smaller value of $\nu_Q$ was observed ($\nu_Q$ ~0. 6 MHz at $x$~0.72).[33, 34]

The longitudinal NMR/NQR relaxation rates divided by $T$, $1/T_1T$, are plotted in Fig. 5 for various values of $x$. The data shown by the open symbols for the aligned samples of $Na_{0.3}CoO_2$ and the powder sample of $Na_{0.3}CoO_2 \cdot 1.3H_2O$, were taken by NQR at the frequency of 3$\nu_Q$. The data shown by the closed symbols for the $Na_{0.3}CoO_2$ sample were taken by NMR at the first satellite line with **H**//**c**. The center transition lines of the aligned samples of $Na_{0.4}CoO_2$ and $Na_{0.6}CoO_2$ were used to take the data shown by the closed symbols with **H**//**c**. Polycrystalline samples were used for $x$~0.5 and 0.75 and the measurements were carried out at the peak positions of the center lines corresponding to the $H$ directions parallel and perpendicular to $c$, respectively. (It should be noted here that in the measurements for $x \sim 0.4$ and $0.5$, the used peaks have the contributions from the two Co sites with $\nu_Q \sim 2.8$ MHz and $4.1$ MHz, and the values shown here should be considered, in some sense, to be the averaged values. Data of $1/T_1T$ measured at each of the two distinct Co sites for $x=0.5$ by zero field NMR are shown later.)

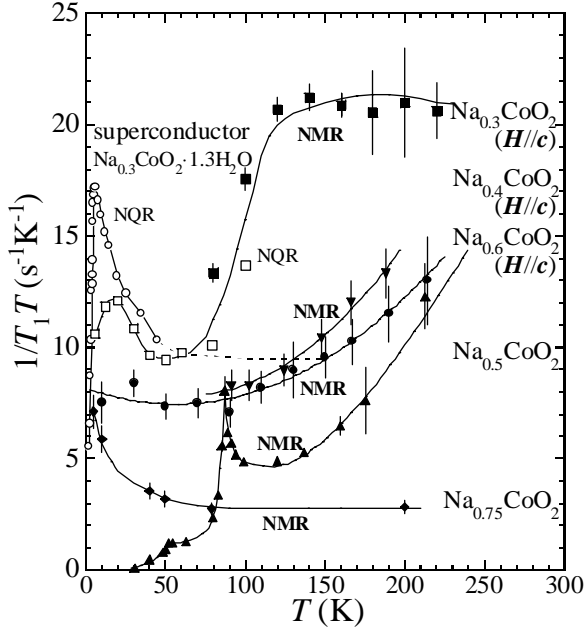

Fig. 5 Longitudinal relaxation rates divided by $T$, $1/T_1T$ measured for powder or $c$-axis aligned samples of $Na_xCoO_2$ are shown as functions of $T$ for various values of $x$. The data of a superconducting sample of $Na_{0.3}CoO_2 \cdot 1.3H_2O$ are also plotted for comparison.

One of the remarkable features of $1/T_1T$ is a rather significant decrease found for $x \leq 0.6$ with decreasing $T$.[34] It reminds us of the pseudo gap phenomenon well known in underdoped Cu oxide high-$T_c$ superconductors.[35,36] For $x \sim 0.75$, it is nearly $T$ independent in the $T$ region above 100K and then increases with decreasing $T$ below 100K. Anomalies due to two phase transitions[28] are observed in the $T$ dependence of $1/T_1T$ for $x=0.5$: The one at $T_{c1}$ exhibits a sharp peak as a function of $T$, and another one at $T_{c2}$ just exhibits a rapid decrease with decreasing $T$.

In the $T$ region below 200 K, $1/T_1T$ of the $Na_xCoO_2$ samples with $x \neq 0.5$ decreases with increasing $x$, which should be contrasted to the increasing tendency of $\chi$ with increasing $x$. For $x \sim 0.3$, $1/T_1T$ is increased by the intercalation of $H_2O$ molecules below $\sim 50$ K. We just note here that the Knight shift does not exhibit significant increase upon this intercalation.

In Fig. 6, the NQR signals at 90 K ($>T_{c1}$) and the zero-field NMR spectra at three temperatures below $T_{c1}$ are shown for a powder sample of $Na_{0.5}CoO_2$. At $T=90.0$ K, NQR peaks corresponding to the frequencies of $3\nu_Q$ at the two distinct Co sites with $\nu_Q \sim 2.8$ MHz and $4.1$ MHz, are observed as indicated by the open and solid arrows, respectively. We have not detected any experimental evidence for nonzero internal magnetic field $H_{int}$. If it exists, the signal should split as observed below $T_{c1}$.

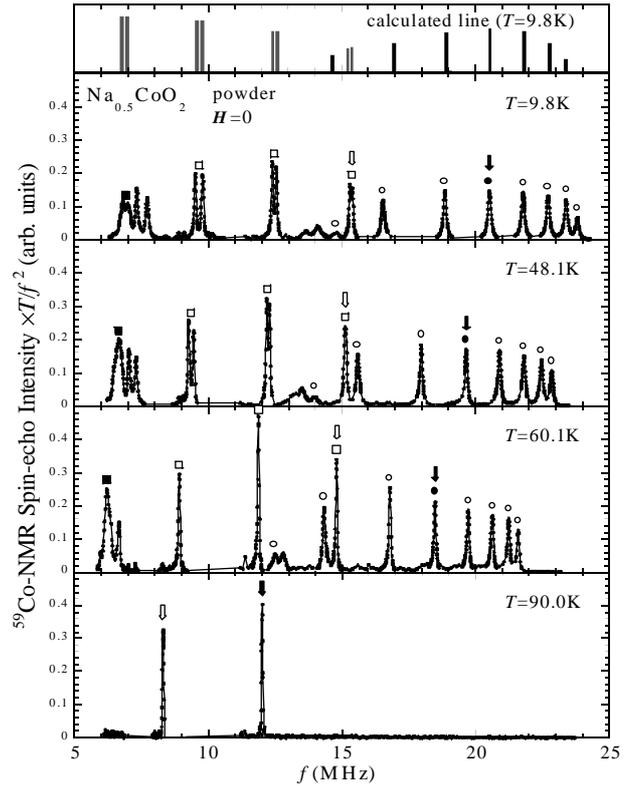

Fig. 6 Data of the NQR and zero field NMR measurements for a powder sample of $Na_{0.5}CoO_2$ are shown, where the spin-echo intensity times $T$ divided by the square of the NMR frequency $f$ are plotted against $f$ at several temperatures $T$. Each of the signal sets indicated by circles and squares exhibits the parallel shift to the higher frequency side below $T_{c1}$ with decreasing $T$. The top panel shows the peak positions and intensities (multiplied by $T$ and divided by $f^2$) calculated for a set by using the parameters $H_{int}$ in $ab$-plane, $H_{int} = 2.0$ T, $\nu_Q = 4.1$ MHz (black bars) and $\eta = 0.3$, and $H_{int}//c$, $H_{int} = 0.7$ T, $\nu_Q = 2.8$ MHz (gray bars). The solid circle and solid square indicates the center transitions. The peaks indicated by the solid and open arrows were used in the $1/T_1T$ measurements, the results of which are shown in Fig. 10 by the open circles and open squares, respectively. The peaks without attached symbol are considered not to be from Co nuclei.

Below $T_{c1}$, we have observed two sets of NMR peaks. These sets can easily be distinguished by their temperature dependences, because the peaks in the same set exhibit a parallel shift to the higher frequency side with decreasing $T$. Analyzing the peak positions of these sets, we have found followings: One set, whose peaks are indicated by the open and closed circles, corresponds to the NMR signals under the internal field $H_{int}$ within the $ab$ plane. It consists of the peaks split by the $eqQ$ interaction with $\nu_Q \sim 4.1$ MHz. $H_{int}$ is $T$ dependent and $\sim 1.8$ T at $\sim 60$ K. (The calculated results of the peak positions and the intensities at 9.8 K for these parameters by considering the $eqQ$ interaction up to the second order are shown in the top panel by the black bars, where $0.25 < \eta < 0.35$.) Another set, whose peaks are indicated by the open and closed squares, corresponds to the signal under the internal field $H_{int}$ parallel to the $c$ axis. It consists of the peaks split by the $eqQ$ interaction with $\nu_Q$ ~2.8 MHz. $H_{int}$ is $T$ dependent and ~0.62 T at 60.1 K. Below $T_{c2}$, the fields split into two values (~0.67 T and ~0.69 T at ~9.8 K). Although the $\nu_Q$ value may also split at $T_{c2}$ with decreasing $T$, we cannot extract the firm conclusion on this point from the present analysis. The peak positions and the intensities obtained at 9.8 K by the similar calculations for the parameters are shown in the top panel by the gray bars.) The solid circles and squares in the figure indicate the central transition lines of these two sets.

We have also found that the values of $H_{int}$ acting on the Co nuclei at two distinct sites have almost the same $T$ dependence. Therefore, the Co moments at these two distinct sites with $\nu_Q$~ 4.1 MHz and the 2.8 MHz are considered to order simultaneously at a temperature $T_{c1}$~87 K with decreasing $T$. The difference between the fields $H_{int}$ at the two distinct Co sites indicates that the Co moments at these sites are different, too. It suggests that the charge order exists at least below $T_{c1}$. Although we do not know at what temperature the charge ordering takes place with decreasing $T$, there exist the two distinct sites even at temperatures above $T_{c1}$, probably higher than 200 K, as is shown later.

(At this point, it should be noted, however, that the fittings of the model calculations are not very precise. In particular, the eight peaks have been observed below $T_{c1}$ for the set with $\nu_Q$~4.1 MHz. The number is larger by one than expected from the line splitting due to the effect of the quadrupole interaction. It is due to the splitting of the highest frequency peak, the origin of which remains unidentified at this moment.)

Figure 7(a) shows the NMR spectra observed for the $c$-axis aligned sample of $Na_{0.5}CoO_2$ with $H$ in the $ab$-plane. In the figure, we can see that as $T$ decreases through $T_{c1}$, about a half of the spectra shifts to both sides of the position above $T_{c1}$, indicating that antiferromagnetic ordering takes place. The magnitude of the observed shift can be explained by the internal field $H_{int}$ (~1.8 T at ~60 K) estimated from the data shown in Fig. 6 for the Co sites with $\nu_Q$~4.1MHz, confirming that the spins at these sites are within the $ab$-plane. The spectra which remain at nearly the same position as observed at 90 K($>T_{c1}$) can be understood as the contribution from the Co sites with $\nu_Q$~2.8 MHz and $H_{int}$(~0.62 T, $H_{int}$//$c$). (The slight shift of the center of gravity of the spectra may be understood by the possible tilting of the moments caused by the external field.)

Figure 7(b) shows the NMR spectra of the aligned crystals of $Na_{0.5}CoO_2$ obtained with $H$//$c$. At 90 K($>T_{c1}$), we see the splitting of the spectrum from the Co sites with $\nu_Q$~2.8 MHz into seven peaks. They are indicated by the red open and closed circles. (Besides these peaks, those which belong to the signals from the sites with $\nu_Q$~4.1MHz also exist and they are indicated by the red arrows.) Below $T_{c1}$, due to the existence of the $T$ dependent internal field $H_{int}$, the peaks from the Co sites with $\nu_Q$~2.8 MHz shift to both the higher and lower fields sides. This can be explained by the existence of antiferromagnetically ordered spins parallel to the $c$ axis. $H_{int}$ estimated here (~0.63 T at ~60 K) is consistent with the value obtained in the analyses of the signals from the $\nu_Q$~2.8 MHz sites shown in Fig. 6. (The peaks from the sites with $\nu_Q$~4.1 MHz are hidden away below $T_{c1}$, because they are broadened upon the magnetic ordering.) Figure 8 shows the temperature dependence of the magnetic susceptibilities $\chi$ measured with $H$//($ab$-plane) and $H$//$c$ for a single crystal sample of $Na_{0.5}CoO_2$. With decreasing $T$, the significant suppression of $\chi$ at $T_{c1}$ is observed only for $H$//($ab$-plane), consistently with the present result that the direction of the larger moments is within the $ab$-plane. With further decreasing $T$, $\chi$ is suppressed in both fields $H$//($ab$-plane) and $H$//$c$ at $T_{c2}$.

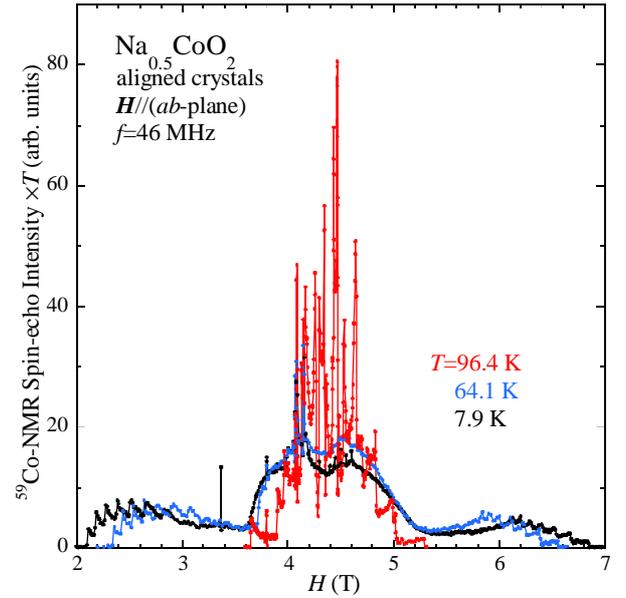

Fig. 7(a) Spin-echo intensity times $T$ measured with the in-plane $H$ for the $c$-axis aligned crystals of $Na_{0.5}CoO_2$ are plotted against $H$ at three temperatures. With decreasing $T$ to ~60 K through $T_{c1}$, about a half amount of signal shifts to both the lower and higher field sides. It indicates that below $T_{c1}$ the internal field exists within the plane due to the antiferromagnetic ordering at one of the two distinct Co sites, that is, the magnetic moments at the sites are in the $ab$-plane. Another half remains near the original position even in the magnetically ordered phase. (The slight shift of the center of the gravity of the spectra is probably due to the tilting of the spins by the external field.) The signal is considered to correspond to the Co sites with magnetic moments along $c$. See the text for details.

Figure 9 shows the $T$-dependence of the internal magnetic field $H_{int}$ obtained from the data shown in Figs. 6, 7(a) and 7(b). NMR absorption profiles observed for a powder sample of $Na_{0.5}CoO_2$ under nonzero applied field (not shown) have also been used to estimate $H_{int}$, where the shift magnitude of the profile front from the position above $T_{c1}$ was considered to be equal to $H_{int}$. The figure indicates that the antiferromagnetic ordering occurs at $T_{c1}$ simultaneously for both Co sites. We have not observed any anomalous behavior at $T_{c2}$ in the $H_{int}$-$T$ curve. These results are different from the reported data of μSR studies.[37,38]

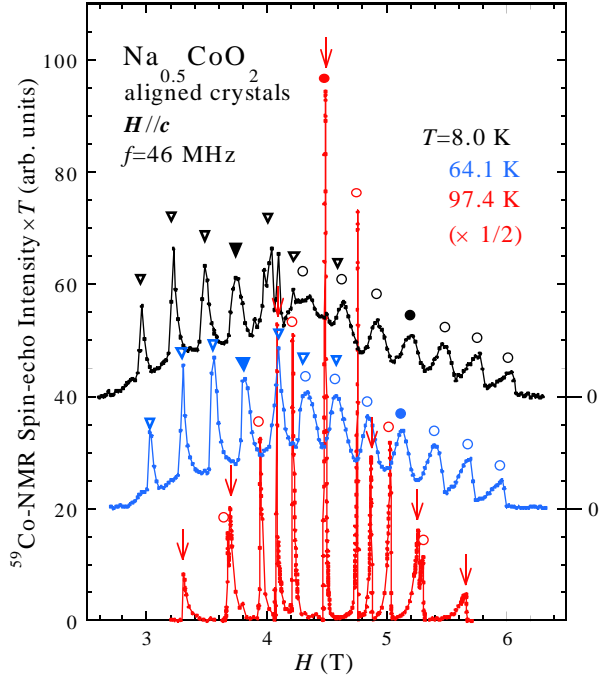

Fig. 7(b) Spin-echo intensity times $T$ measured with $H//c$ for the $c$-axis aligned crystals of $Na_{0.5}CoO_2$ are plotted against $H$ at three temperatures. A set of peaks split by the $eqQ$ interaction as indicated by the circles at ~90 K (>$T_{c1}$), exhibits a shift to both the lower and higher field sides as shown by the circles and triangles, respectively, when the internal magnetic field increases with decreasing $T$ below $T_{c1}$. These peaks correspond to the $\nu_Q$ value of ~2.8 MHz or the smaller magnetic moment. (The closed circles and triangles correspond to the center transitions.) The data indicate that the moments are along the $c$-axis. Another set of the peaks from the Co sites with $\nu_Q$~4.1 MHz are indicated by the arrows. These peaks are from the sites with $\nu_Q$~4.1 MHz and hidden away below $T_{c1}$, because they are broadened by the internal field within the plane.

Figure 10 shows the $T$-dependences of $1/T_1T$ of the powder sample of $Na_{0.5}CoO_2$, where the open circles and open squares show the results obtained at the peak positions indicated by the closed and open arrows in Fig.6, respectively. The closed circles present the same data shown in Fig. 5, which were obtained for the centerlines of the NMR spectra contributed from both of two distinct Co sites. The $1/T_1T$ values observed at the sites with $\nu_Q$~4.1MHz (open circles) are about three times larger than those with $\nu_Q$~2.8 MHz (open squares) at temperatures $T>T_{c1}$, indicating that the spin fluctuations are stronger at the Co sites with $\nu_Q$~4.1MHz than those at the sites with $\nu_Q$~2.8MHz. One might expect that even above $T_{c1}$, there exists the charge disproportionation which induces the difference of the magnetic fluctuation strengths between the distinct sites. However, the observed difference of the relaxation rates does not necessarily indicate the existence of the charge order above $T_{c1}$.

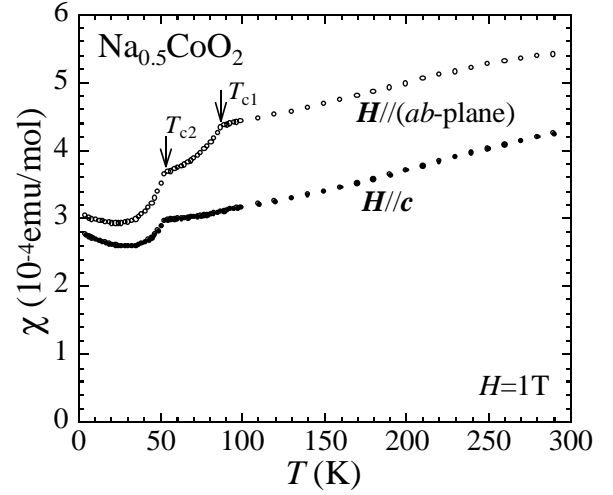

Fig. 8 Magnetic susceptibilities measured for a single crystal of $Na_{0.5}CoO_2$ with the field of 1 T parallel and perpendicular to $c$ are shown. At $T_{c1}$, the anomaly is very small for $H//c$, while the anomalies are significant at $T_{c2}$ for both directions of $H$ perpendicular and parallel to the $c$-axis.

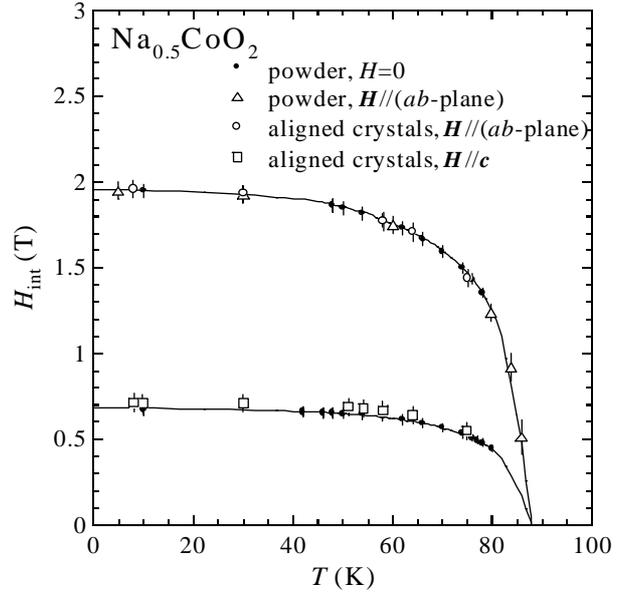

Fig. 9 Internal magnetic fields $H_{int}$ observed for powder or $c$-axis aligned samples of $Na_{0.5}CoO_2$ at the two distinct Co sites are shown against $T$. The fields appear at $T_{c1}$ and no appreciable anomaly has been detected in their $T$ dependence at $T_{c2}$.

The sharp peaks observed at $T_{c1}$ in the $1/T_1T$-$T$ curves for both kinds of Co sites indicate that the antiferromagnetic ordering takes place at the temperature with decreasing $T$. The $1/T_1T$ measured for the overlapped centerlines of the two Co sites are almost equal to that of the sites with $\nu_Q$~2.8 MHz. It is due to a fact that the $1/T_1T$ measurements at the centerlines see the values of the slower component of the two. With decreasing $T$ from $T_{c1}$ to $T_{c2}$, the values of $1/T_1T$ at the two distinct sites decrease rapidly and seem to become almost equal to each other's at $T_{c2}$.

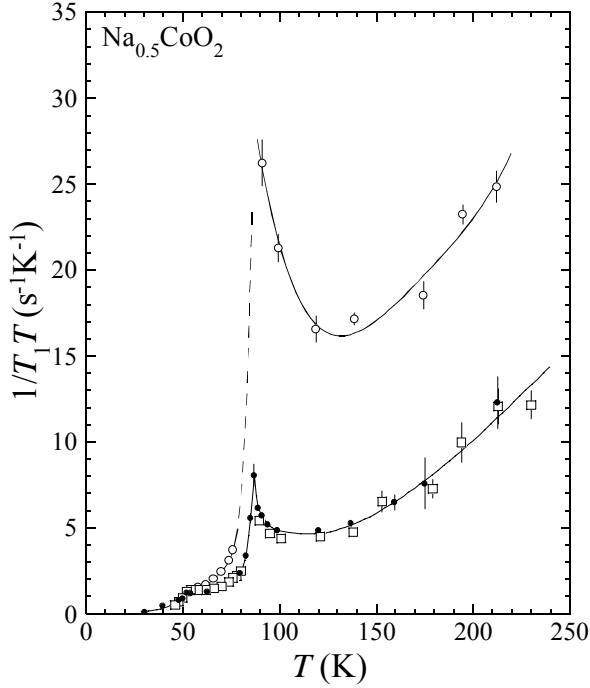

Fig. 10 The NMR longitudinal relaxation rates divided by $T$, $1/T_1T$ measured for the powder sample of $Na_{0.5}CoO_2$. The open circles and open squares shows the results obtained at the peak positions indicated by the closed and open arrows in Fig.6, respectively. The closed circles present the same data shown in Fig. 5, which were obtained for the centerlines of the NMR spectra contributed from both of two distinct Co sites.

*3.3 Neutron measurements*

In *3.2*, we have presented NMR/NQR data on $Na_{0.5}CoO_2$, where following facts have been revealed. (1)The magnetic ordering takes place at $T_{c1}$ with decreasing $T$. (2)There are two distinct Co sites with different magnetic moments $\mu_1$ and $\mu_2$ ($\mu_1 > \mu_2$), where $\mu_1$ is within the *ab* plane and $\mu_2$ is parallel to the *c*-axis. (3)The charge order exists at least below $T_{c1}$, and so on.

Now, the next step of the study is to clarify the ordering pattern of the magnetic moments in $Na_{0.5}CoO_2$. Here, we have searched magnetic reflections for the two crystals of $Na_{0.5}CoO_2$ as stated above and observed them at the superlattice points $Q=(h,h,l)$ and $(h,0,l)$ in the reciprocal space with half integer $h$ and odd $l$. In Fig. 11(a), the observed profiles of 1/2 1/2 1 are shown at 15 K ($<T_{c2}$), 60 K (between $T_{c2}$ and $T_{c1}$) and 120 K ($>T_{c1}$). The integrated intensity of this reflection is plotted against $T$ in Fig. 11(b). Analyzing the integrated intensities of the superlattice above $T_{c1}$, we have found that they can be well understood by the Na ordering pattern proposed in refs. 39 and 40. Then, for all superlattice reflections, the intensities observed above $T_{c1}$

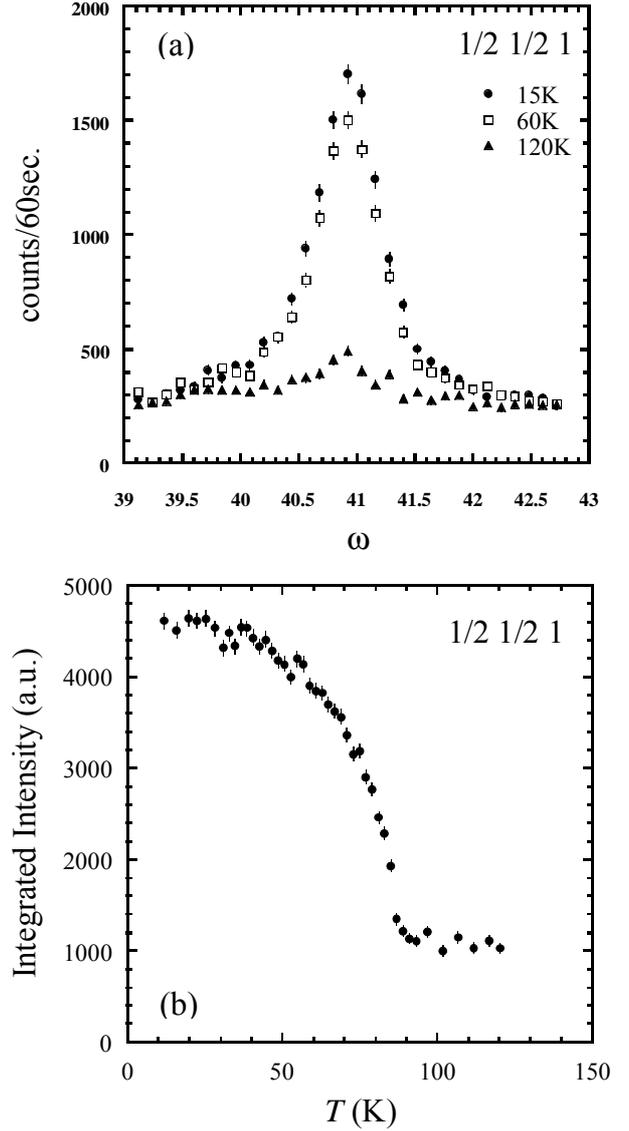

Fig. 11 (a) The profiles of ω-scans of 1/2 1/2 1 reflection at 15K, 60K and 120K. (b) $T$-dependence of the integrated intensities of the 1/2 1/2 1 reflection between 15K and 120K.

were subtracted from those below $T_{c1}$. The resulting data ($I_{obs}$) were fitted by the calculated values ($I_{cal}$) of magnetic structure models, where the distribution of the magnetic domain was considered as described in Appendix A. Before discussing the results of the fittings, we note that the magnetic ordering takes place at $T_{c1}$ with decreasing $T$, which is consistent with the NMR/NQR result. We also note that we do not observe any appreciable anomaly at $T_{c2}$ within the experimental accuracy, while the zero-field NMR observes the splitting of the peaks for the signals from the Co sites with $\nu_Q \sim 2.8$ MHz.

Figures 12(a) and 12(b) show the results of the fittings, $I_{obs}$ *versus* $I_{cal}$ and the magnetic ordering patterns used in the model calculations. In tables 1(a) and 1(b), their numerical values are also compared. At this moment, we cannot distinguish which one of the two patterns is more realistic. It is because the differences of the intensities between the superlattice reflections, which can be used to distinguish the two magnetic structures, cannot be obtained so precisely for the reasons that $\mu_2$ is small and that there exist the contributions due to the Na ordering at the superlattice points. In the calculations, we have considered that the larger moments ($\mu_1$) is within the *ab* plane and the smaller one ($\mu_2$) is perpendicular to the plane and we have assumed the approximate relation $\mu_1/\mu_2=3$, which is rationalized for the reasons described in the Appendix B.

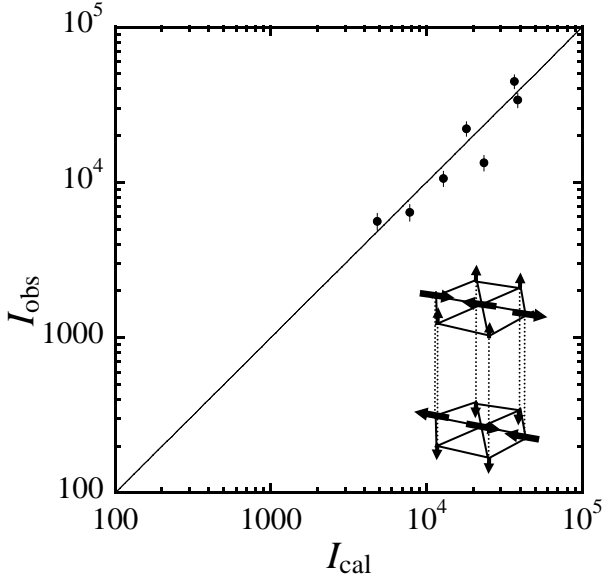

Fig. 12(a)  Results of the fitting to the observed integrated intensities of the magnetic reflections by the calculation for the magnetic structure model shown in the figure. The value of $\mu_1=0.34\pm0.03$ $\mu_B$ ($\mu_2$ was fixed at $\mu_1/3$).

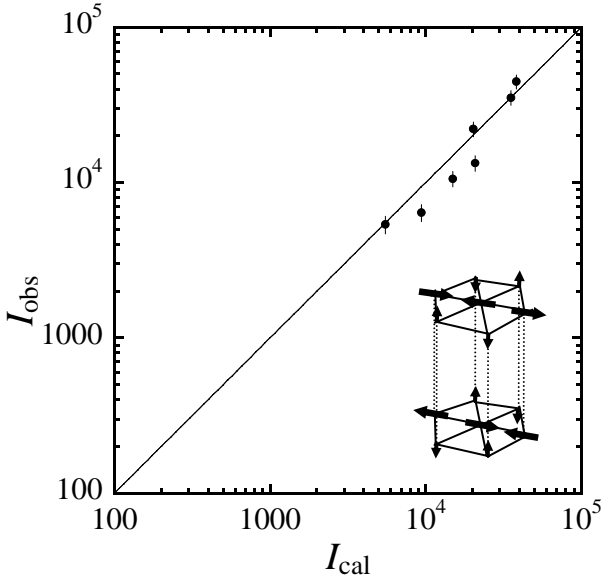

Fig. 12(b)  Results of the fitting to the observed integrated intensities of the magnetic reflections by the calculation for the magnetic structure model shown in the figure. The value of $\mu_1=0.34\pm0.03$ $\mu_B$ ($\mu_2$ was fixed at $\mu_1/3$).

Table 1. Observed($I_{obs}$) and calculated($I_{cal}$) values of the integrated intensities of the magnetic reflections are compared. The magnetic intensities are estimated by subtracting the values at 120 K from those at 8 K. The results in (a) and (b) correspond to the magnetic structures shown Figs. 12 (a) and (b), respectively.

(a)

|         | $I_{obs}$     | $I_{cal}$ |
|---------|---------------|-----------|
| 1/21/21 | 22218 ± 2315  | 18080     |
| 1/21/23 | 10585 ± 1164  | 12856     |
| 1/21/25 | 6401 ± 772    | 7802      |
| 1/201   | 44676 ± 4521  | 36724     |
| 1/203   | 35283 ± 3582  | 38513     |
| 1/205   | 13355 ± 1465  | 23363     |
| 3/201   | 5375 ± 676    | 4824      |

(b)

|         | $I_{obs}$     | $I_{cal}$ |
|---------|---------------|-----------|
| 1/21/21 | 22218 ± 2315  | 20276     |
| 1/21/23 | 10585 ± 1164  | 14994     |
| 1/21/25 | 6401 ± 772    | 9379      |
| 1/201   | 44676 ± 4521  | 38441     |
| 1/203   | 35283 ± 3582  | 35352     |
| 1/205   | 13355 ± 1465  | 20898     |
| 3/201   | 5375 ± 676    | 5524      |

In both cases of Figs. 12(a) and 12(b), we have obtained reasonable results for $\mu_1\sim0.34\pm0.03\mu_B$ and $\mu_2=\mu_1/3$, which guarantees that the intensity-components, which additionally appear at the superlattice points at $T_{c1}$ with decreasing $T$, are magnetic.

Now, we have presented possible magnetic structures, which can explain both the neutron and NMR/NQR results. The observation of the two different magnetic moments suggests the existence of the charge ordering of the Co sites into $Co^{3.5+\delta}$ and $Co^{3.5-\delta}$ and one of the interesting points is that the ordering pattern is closely related to that of Na atoms (see Fig. 13).[39]

**4. Discussion**

As shown in Figs. 1 and 5, the $T$ dependences of $\chi$ and $1/(T_1T)$ of $^{59}$Co nuclei of Na$_x$CoO$_2$ exhibit quite different behaviors between the regions divided by $x_c=0.6$. On this point, Cava's group[28] have pointed out that $\chi$ has the Curie-Weiss component in the region of $x>0.5$, while it is Pauli paramagnetic for $x<0.5$. Here, we emphasize that in the region of $x\leq0.6$, both of $\chi$ and $1/T_1T$ $^{59}$Co nuclei exhibit the quite characteristic $T$ dependence reminiscent of the well known pseudo-gap behavior of high-$T_c$ Cu oxides: The magnetic susceptibility $\chi$ has a peak near room temperature and decreases with decreasing $T$ and $1/T_1T$ also exbibit the similar $T$ dependence. This tendency implies that there exists an antiferromagnetic correlation as in underdoped high-$T_c$ Cu oxides.

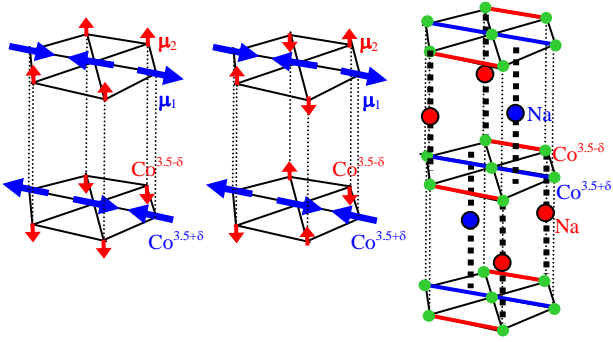

Fig. 13 The magnetic structures proposed here for $Na_{0.5}CoO_2$ are shown with the ordered pattern of Na atoms. The close relationship between the ordering patterns of the Na atoms and the Co moments can be seen.

NMR spectra of $Na_xCoO_2$ depend on the $x$ value. In the region of $x \geq 0.7$, the signal with $\nu_Q \sim 0.6$ MHz (not shown) was observed. For the samples with $x \leq 0.6$, signals from two kinds of Co sites with the $\nu_Q$ values of ~2.8 and ~4.1 MHz were observed, where the latter site becomes dominant at $Na_{0.3}CoO_2$ as shown in Fig. 4. Because only the Co sites with $\nu_Q = \sim 4$ MHz have been observed in the superconducting $Na_{0.3}CoO_2 \cdot 1.3H_2O$,[19] the Co sites with $\nu_Q = \sim 4$ MHz may become important for the discussion of the physical properties related to the superconductivity.

Now, in the ordered phase of $Na_{0.5}CoO_2$ ($T < 87$ K), the moment values $\mu_1$ and $\mu_2$ at the two Co sites with $\nu_Q$ ~4.1 and 2.8 MHz, respectively, can be estimated from the neutron diffraction data to be $\mu_1$ ~0.34±0.03 $\mu_B$ and $\mu_2$ ~$\mu_1/3$ at 9.8 K. The possible magnetic structure is either one of those shown in Figs. 12(a) and 12(b). The ordered moments at these two sites have the almost same $T$ dependences below 87 K and the patterns are closely related to that of Na atoms. These facts indicate that these sites exist as a microscopically homogeneous phase, not as the phase-separated one. The close relationship between the ordering patterns of the two distinct Co sites with that of Na ions[39] indicates that for the determination of the electronic structures and the magnetic moments of Co atoms, the arrangement of the Na sites between the $CoO_2$ planes seems to be important: As shown in Fig. 13, chains of $Co^{3.5+\delta}$ and $Co^{3.5-\delta}$ exist alternatingly within the $CoO_2$,[39,40] and one of the crystallographically distinct sites of Na ions is located above or below the Co sites of the $Co^{3.5-\delta}$ chain in an alternating way. Another one is located above or below the center of gravity of the Co triangles formed by $Co^{3.5+\delta}$ and $Co^{3.5-\delta}$. (We do not think that the former Na sites are located above or below the Co sites of the $Co^{3.5-\delta}$. It is because because the Coulomb interaction energy is of the disadvantageous.) Because $Co^{3.5+\delta}$ ions are expected to have a larger moment, we can say that the Co-sites with the larger $\nu_Q$ and larger ordered moment within the $ab$-plane, have the valency of $3.5+\delta$, while those with the smaller $\nu_Q$ and smaller ordered moment parallel to $c$-axis correspond to the valency of $3.5-\delta$.

Results of neutron scattering studies on $Na_{0.82}CoO_2$[25] and $Na_{0.75}CoO_2$[26,27] have pointed out followings. The in-plane and inter-plane spin correlation is ferromagnetic and antiferromagnetic, respectively, and the spins are along the $c$-axis. The inter plane exchange interaction of the samples with $x > 0.7$ is comparable with that of in-plane one, resulting in the three dimensional nature of the spins system. It is somewhat unexpected from the layered structural characteristic of the system. The origin of this three dimensional nature has been argued by considering the inter plane Co-Co interaction via Na sites (the $sp^2$ hybridized orbit).[41] Then, what things are expected to take place for the dimensionality and the spin-spin correlation, when the Na content $x$ decreases. First, the 2-dimensional nature is enhanced because of the decreasing number of the $sp^2$ hybridized orbits. Considering the possible spin structures of $Na_{0.5}CoO_2$ shown in Figs. 12(a) and 12(b), we also expect that the main in-plane spin correlation changes to the antiferromagnetic one. (At least, the spin correlation of the larger moment-sites, which do not have Na ions above or below their sites as shown in Fig. 13, are antiferromagnetic. Although we cannot distinguish, as we stated above, if the in-plane spin correlation of the smaller moment sites is ferromagnetic or antiferromagnetic at $x=0.5$, we speculate that the antiferromagnetic correlation becomes dominant with decreasing $x$ to 0.3, because the occupation of Na atoms above or below their sites becomes smaller.) Then, it is quite plausible that the two dimensional electrons with antiferromagnetic correlation is realized in the region of small $x$ values for $Na_xCoO_2$. The intercalation of $H_2O$ molecules enhances the two dimensional nature of the electrons. These considerations support the pseudo-gap-like behavior observed in the measurements of $\chi$ and $1/T_1T$, shown in Figs. 1 and 5, respectively.

Now, the present results imply that the hydrated system has the singlet Cooper pairs, which was reported by the present authors' group based on the Knight shift data.[22,23] For the occurrence of the triplet pairing, we have to expect that the intercalation of $H_2O$ molecules induces the change of the Fermi surface shape or its topology,[42] which brings about the drastic change of the magnetic correlation in the system from antiferromagnetic to ferromagnetic one. However, it may not be realistic, because we have not observed a tendency of the ferromagnetic correlation in the $T$ dependence of the Knight shift.[22] In any case, it seems to be important for the complete understanding of the superconductivity of $Na_{0.3}CoO_2 \cdot yH_2O$, to construct a consistent explanation of various kinds of experimental results.[12-23]

## 5. Summary and Conclusions

Data of various kinds of measurements have been presented for $Na_xCoO_2$. In particular, magnetic properties and magnetic structures of $Na_{0.5}CoO_2$ are reported in detail. Attentions have also been paid to the $x$ dependence of the electronic nature. There exists a rather clear phase boundary at $x=x_c\sim0.6$ and for $x<x_c$, the pseudo-gap-like behavior can be observed in the temperature dependence of the magnetic susceptibility and the NMR/NQR $1/T_1T$.

For $Na_xCoO_2$ in the region of $x>0.7$, the NMR signal with $\nu_Q\sim0.6$ MHz was observed. For the samples with $x<0.6$, signals from two kinds of Co sites with different $\nu_Q$ values, ~4.1 MHz and ~2.8 MHz were observed, and signals from one of these sites with $\nu_Q\sim4.1$ MHz become dominant with decreasing $x$ to 0.3. Because only the Co sites with $\nu_Q\sim4$ MHz have been observed in the superconducting $Na_{0.3}CoO_2\cdot1.3H_2O$,[19] the Co sites with $\nu_Q\sim4$ MHz may become important for the discussion of the physical properties related to the superconductivity.

We have pointed out that in $Na_{0.5}CoO_2$, the Co sites with this larger $\nu_Q$ value have the larger magnetic moments. These moments align antiferromagnetically at $T_{c1}\sim87$ K with their direction within the plane, while other Co sites with the smaller $\nu_Q$ value have the smaller moments. They align with the moment direction parallel to the $c$ axis. The ordered moments of the two Co sites exhibit the same $T$ dependences, indicating that the existence of the two distinct sites is not due to a macroscopic phase separation. Although we have observed at $T_{c2}$ the splitting of the zero-field NMR signals from the Co sites with the $\nu_Q\sim2.8$ MHz, any anomalous $T$ dependence of the ordered magnetization has not been observed, at $T_{c2}\sim53$ K, in the $T$ dependences of the intensities of the neutron magnetic reflections. At least, the transition at $T_{c2}$ does not change the magnetic structure drastically. Then, a question arises why the anomalies observed at $T_{c2}$ in the $T$ dependence of $\chi$ shown in Fig. 8 is so significant. It remains as a future problem.

As the Na concentration $x$ is decreased, the two dimensional nature of the electrons of $Na_xCoO_2$ is enhanced and the antiferromagnetic correlation increases. These are due to the decrease of the occupancy of Na atoms, which have a bridging role of the inter layer Co-Co exchange interaction. Even if the magnetic interaction is important for the occurrence of the superconductivity, it seems to be the antiferromagnetic ones and therefore the Cooper pairs are expected to be in the singlet state. It is consistent with our recent results of the Knight shift studies.

Difference of the ordered moments between the two distinct Co sites below $T_{c1}$ indicates that the charge ordering exists at least below this temperature. Above $T_{c1}$, we have found that these two kinds of Co sites have the different relaxation rate $1/T_1$. However, it does not indicate that there exists the charge disproportionation, but may just indicate the difference of the strength of the spin fluctuation between the two distinct sites. We do not know if the charge ordering exists even above $T_{c1}$. We have not clarified what kind of change takes place in the electronic state at $T_{c2}$, either. These remain as the future problems.

**Acknowledgements**

Work at JRR-3M was performed within the frame of JAERI Collaborative Research Program on Neutron Scattering. The work is supported by Grants-in-Aid for Scientific Research from the Japan Society for the Promotion of Science (JSPS) and by Grants-in-Aid on priority area from the Ministry of Education, Culture, Sports, Science and Technology.

**Appendix A**

The distribution of the magnetic domains are considered in the fittings of the calculated intensities to the observed ones. Considering that the $c$-axis and one of the in-plane axis along the Co-Co direction(say $a$-axis) of $Na_{0.5}CoO_2$ is perpendicular to the crystal rod prepared by the floating zone method, the probability that the chains of the $\mu_1$ sites run along this $a$-axis can be different by a factor of $p$ from two other in-plane axes. Then, $p$=4.5 and 2.5 were obtained for the magnetic structures shown in Figs. 12(a) and 12(b), respectively. It can roughly explain the integrated intensities of the superlattice reflections due to the Na ordering observed above $T_{c1}$, too.

**Appendix B**

For the model structure shown in Fig. 12(b), the transferred hyperfine interaction does not exist between the $\mu_1$ and $\mu_2$ sites, and therefore the sum of the contributions of the on-site-hyperfine field($A_1\mu_1$) and the transferred-hyperfine field ($B_1\mu_1$) between the $\mu_1$ sites, $(A_1+2B_1)\mu_1$ determines the hyperfine field at the $\mu_1$ sites, while the field at the $\mu_2$ sites is determined by $(A_2+2B_2)\mu_2$ with the similar notations of the hyperfine couplings. Then, assuming the relation $(A_1+2B_1)\sim(A_2+2B_2)$, we use the approximation that $\mu_1/\mu_2$ is equal to the ratio (~3) of the hyperfine fields observed by the NMR/NQR studies at the $\mu_1$ and $\mu_2$ sites. If the in-plane ordering pattern of the moments $\mu_2$ is ferromagnetic as shown in Fig. 12(a), the total hyperfine fields at the $\mu_1$ and $\mu_2$ sites are described as $[((A_1+2B_1)^2\mu_1^2+(4B_2)^2\mu_2^2)]^{1/2}$ and $(A_2+2B_2)\mu_2$. If $\mu_2^2$ is much smaller than $\mu_1^2$, we can use the approximate relation $\mu_1/\mu_2\sim3$. If $\mu_2$ is as large as $\mu_1$, we cannot neglect the second term of the square bracket and the ratio $\mu_1/\mu_2$ may not be approximated by the ratio of the hyperfine fields. In the present case, we can find that $\mu_2$ is not as large as $\mu_1$ from the experimental result that the 111 reflection intensity, which is contributed from the $\mu_2$ moments, is much smaller than the value expected for $\mu_2=\mu_1$, confirming that $\mu_1/\mu_2$ can be approximated by the ratio of the hyperfine fields. (We cannot fully exclude, however, the possible existence of $\mu_2\sim\mu_1/3$ due to the large experimental error bar for the intensity of the 111 magnetic reflection, because at the Bragg point, there exists another contribution due to the Na ordering. It is why we cannot distinguish which one of the two magnetic structures shown in Fig. 12(a) and

12(b) is actually realized.)